\documentstyle[amssymb,aps,prl,preprint]{revtex}

\begin{document}
\title{Density Matrix Renormalization Group Study of One-Dimensional Acoustic
Phonons.}
\author{L.G. Caron and S. Moukouri}
\address{Centre de recherche en physique du solide and D\'epartement de physique,\\
Universit\'e de Sherbrooke, Sherbrooke, QC, Canada J1K 2R1.}
\date{26 May 1997}
\maketitle

\begin{abstract}
We study the application of the density matrix renormalization group (DMRG)
to systems with one-dimensional acoustic phonons. We show how the use of a
local oscillator basis circumvents the difficulties with the long-range
interactions generated in real space using the normal phonon basis. When
applied to a harmonic atomic chain, we find excellent agreement with the
exact solution even when using a modest number of oscillator and block
states (a few times ten). We discuss the use of this algorithm in more
complex cases and point out its value when other techniques are deficient.
\end{abstract}


In a previous paper\cite{PRL}, we have studied how the real-space density
matrix renormalization group (DMRG) \cite{Noack}\cite{White} could be
applied to {\it dispersionless} phonons in the ground state of magnetic or
fermion chains. We showed that the unavoidable truncation of the local
oscillator space proved manageable for a Peierls or spin-Peierls
deformation. Acceptable numerical accuracy could be obtained by properly
selecting the dimension of the oscillator space as a function of the
amplitude of the lattice deformation (the Peierls gap).

The true Peierls, spin-Peierls, and superconductivity problems, or more
generally the electron-phonon problem, however, involve acoustic phonons
having {\it dispersion}. Are the hydrodynamic modes problematic? Let us look
at other numerical approaches and see how they deal with phonons. Although
the dispersion has not created any difficulty in exact diagonalization{\it \ 
}calculations on a one-dimensional (1D) lattice \cite{Wang}, most efforts
with electron-phonon systems have used dispersionless phonons \cite
{dispersionless} or have focused on a single acoustic mode related to an
order parameter \cite{single mode}. The reason was to keep the dimension of
the quantum space tractable. The chain or cluster size being rather small
(of order 10 sites), however, there is little hope of doing numerical
justice to the hydrodynamic modes. The quantum Monte Carlo method (QMC) \cite
{QMC} has apparently done very well with acoustic phonons and the
hydrodynamic modes. The QMC does have sign problems, however, with the
electronic part in frustrated systems and with some magnetic impurity
problems \cite{Zhang}. There is no problem with phonons, of course, in a
momentum space formulation of the DMRG \cite{Xiang}. The treatment of local
interactions (Hubbard interaction, exchange interaction, electron-phonon
interaction...) is, however, more complex and computationally exacting in
momentum space.

The DMRG has the capability of reaching chain lengths in the hundreds, is
free of the sign problem of the QMC, and is ideally suited to short-range
interactions. The situation with the hydrodynamic modes is {\it potentially}
troublesome, however. The long-range displacement-displacement
autocorrelation function $\left\langle (u_\ell -u_{\ell ^{\prime
}})^2\right\rangle $ in an infinite chain increases logarithmically with the
distance $R_{\ell \ell ^{\prime }}$ between the static (classical) atomic
equilibrium positions. The same is true of the displacement amplitude
squared $\left\langle (u_\ell )^2\right\rangle $ at the center of a finite
chain with closed boundary conditions (fixed ends) which grows
logarithmically with the number $N_a$ of atoms in the chain. As a
consequence, the amplitude of the atomic displacements can grow without
bounds as the chain length increases in the DMRG procedure. In other words,
there is no fixed point for the infinite-chain DMRG algorithm. It is forever
fleeting as the chain grows. There is also another constraint on the form
the Hamiltonian should have in the DMRG. The procedure takes place in real
space and, ideally, the Hamiltonian should be reducible to a form involving
only short-range coupling terms between the central site(s) and the two
adjacent blocks. Otherwise, the numerical accuracy and memory expenditures -
for the storage of the block variables coupling to the central site(s) -
suffer greatly. It so turns out, as we shall see below, that using the
normal phonon coordinates, defined in momentum space, leads to long-range
couplings in real space and thus to potential difficulties for the DMRG.

It is the purpose of this paper to study the constraints imposed by these
considerations on the DMRG treatment of the ground state of a chain bearing
acoustic phonons. We have chosen to focus specifically on the phonons
leaving aside any electronic or spin counterpart. There are well documented
applications of the DMRG to spin \cite{White}\cite{spin}\cite{Moukouri},
electron \cite{electron} or mixed \cite{mixed} systems. As such, the atomic
chain problem is more a testing ground for the DMRG than a physical problem
since its excitations and thermodynamics have exact solutions.

We shall study finite chains having $N_a$ atoms and closed boundary
conditions. Keeping the end atoms at fixed positions eliminates the
troublesome $q=0$ mode which, for periodic or open boundary conditions,
corresponds to uniform displacements (and unbounded $\left\langle (u_\ell
)^2\right\rangle $). We shall assume the usual harmonic Hamiltonian 
\begin{equation}
H=\sum_\ell \frac{p_\ell ^2}{2m}+\frac 12K\sum_\ell \left( u_\ell -u_{\ell
+1}\right) ^2  \label{H0}
\end{equation}
and the boundary conditions $u_1=u_{N_s}=0$. The solutions are of course
well known in terms of the annihilation $d_q$ and creation $d_q^{\dagger }$
operators for the normal modes of momentum $q=n\pi N_{mod}^{-1}$, $1\leq
n\leq N_{mod}$, where $N_{mod}=(N_a-2)$ is the number of normal modes. One
has simply 
\begin{equation}
H=\sum_q\hbar \omega _q(d_q^{\dagger }d_q+\frac 12)\ ,  \label{H1}
\end{equation}
where the eigenfrequency is $\omega _q=2\sqrt{K/m}\left| \sin (q/2)\right| $%
. Here and throughout, we have put the static equilibrium interatomic
distance equal to one.

The need for writing the Hamiltonian in real space stems from the real space
algorithm used in the DMRG. In the case of our Eq. \ref{H1}, one could try
to exploit the operators $d_\ell =N_{mod}^{-\frac 12}\sum_q\exp (iq\ell )\
d_q$ on the perhaps natural reflex of using the diagonal phonon annihilation
operators. Unfortunately, the Hamiltonian in real space then has the form 
\begin{eqnarray}
H &=&\sum_{\ell ,\ell ^{\prime }}F(\ell ^{\prime }-\ell )d_{\ell ^{\prime
}}^{\dagger }d_\ell +\frac 12\sum_q\hslash \omega _q\ ,  \label{H2} \\
&&F(x)=\ N_{mod}^{-1}\sum_q\hslash \omega _q\exp (iqx)\ .  \nonumber
\end{eqnarray}
The $F(x)$ function is slowly decreasing, varying as $x^{-2}$, at long
distances. Consequently, the block operators ${\cal O}_{b\ell ^{\prime
}}=F(\ell ^{\prime }-\ell )d_{\ell ^{\prime }}$ are relevant for all sites $%
\ell ^{\prime }$ in the blocks. The storage requirements are large. The
number of stored elements is of order $N_{bs}M_b^2$ at each step $s$, for
the infinite-chain algorithm, where $N_{bs}$ and $M_b$ are the number of
atoms and the number of selected states in each block, respectively. It is
huge, of order $\sum_sN_aM_b^2$, for the finite-chain algorithm, the sum
covering all steps $s$ of the procedure. The numerical accuracy would
greatly suffer.

There is a second problem, for the infinite-chain algorithm, related to the
quantization of the phonon momentum $q$. Its values depend on the chain
length $N_a$ and thus change from step to step. Consequently, so does the
amplitude $F(x)$ of the coupling terms. This is rather annoying. This
problem can of course be circumvented by using the finite-chain algorithm as
one can use the $q$ values of the chosen chain length.

The way around these problems is to use another basis set of quantum
oscillators. Let us use instead the local oscillators (as in \cite{Wang})
that are solutions of 
\begin{equation}
H_o=\sum_\ell \frac{p_\ell ^2}{2m}+K\sum_\ell \left( u_\ell \right)
^2=\sum_\ell \hslash \omega _o(b_\ell ^{\dagger }b_\ell +\frac 12)\ ,
\label{H3}
\end{equation}
where $\omega _o=\sqrt{2K/m}$, $u_\ell =\left( \frac{\hslash }{2m\omega _o}%
\right) ^{\frac 12}(b_\ell ^{\dagger }+b_\ell )$. With this construction,
one can write 
\begin{equation}
H=H_o+H_c\ ,  \label{H4}
\end{equation}
where 
\[
\quad H_c=-K\sum_\ell u_\ell u_{\ell +1}=-\frac{\hslash \omega _o}4\sum_\ell
(b_\ell ^{\dagger }+b_\ell )(b_{\ell +1}^{\dagger }+b_{\ell +1})\ . 
\]
This expression for $H$ is exact. The coupling terms $H_c$ are now, however,
short ranged and independent of the chain length. This last formulation is
thus highly preferable for use with the DMRG even though the local
oscillators are not diagonal. As a matter of fact, the connection between
the $b_\ell $ and the $d_q$ is not trivial: 
\begin{eqnarray}
b_\ell &=&N_{mod}^{-\frac 12}\sum_q\left[ (\omega _o/2\omega _q)^{\frac 12%
}+(\omega _q/2\omega _o)^{\frac 12}\right] \sin (q\ell )d_q  \label{bd} \\
&&+\left[ (\omega _o/2\omega _q)^{\frac 12}-(\omega _q/2\omega _o)^{\frac 12%
}\right] \sin (q\ell )d_q^{\dagger }\ .  \nonumber
\end{eqnarray}
\newline
This is a canonical transformation. We shall now study the implementation of
Eq. \ref{H4}.

As mentioned in the Introduction, the central atom(s) oscillator space must
be truncated at a dimension sufficient to properly represent the vibrational
motion of the atoms. Regardless of the algorithm used, one must begin by
using the infinite-chain algorithm which is plagued by the absence of any
fixed point. Indeed, the vibrational amplitude of the central atom was
already mentioned to be logarithmically increasing at each step: 
\begin{equation}
\left\langle (u_\ell )^2\right\rangle =N_{mod}^{-1}\sum_q\left( \frac{%
\hslash }{m\omega _q}\right) \left( \sin (q\ell )\right) ^2\varpropto \ln
(N_a)\ .  \label{u2}
\end{equation}
This last quantity is a measure of the average occupation number $%
\left\langle n_\ell \right\rangle $ of the local oscillator. It is then a
prerequisite that the dimension $M_\nu $ of the local oscillator space be
sufficient to cover the requirements of Eq. \ref{u2}, that is $M_\nu \gg
\left\langle n_\ell \right\rangle $. Our previous experience\cite{PRL} has
revealed that this dimension should be of order ten. We have thus used only
one central atom (site) in order to keep the computation time and storage
requirements within acceptable values. We have monitored the occupation
probability $P_s(n)$ of each oscillator state $|n\rangle =(n!)^{-\frac 12%
}(b_\ell ^{\dagger })^n|0\rangle $, $0\leq n\leq M_\nu $, of the central
atom for each step $s$ of the infinite-chain algorithm for $N_a$ up to 49
atoms. We find that $P_s(n)\approx 0.35\exp (-n/\alpha _s)$, $\alpha
_s\approx .33+.18\ln (N_a-3)$, fits the data quite well for most
occupations, except for the first and last ones $n=0,M_\nu -1$ . Notice the
logarithmic dependence for $\alpha _s$ which weighs the average and the
standard deviation for the local oscillator occupation. Having this
information, one can estimate a minimal relative error on the oscillator
space statistics $\delta P_s\approx P_s(M_\nu -1)$ at each step resulting
from using a finite dimensional oscillator space for the central atom. This
error is propagative and thus, the total minimum numerical error of a sweep
should be expected to be of order $\delta P\approx \sum_sP_s(M_\nu )$. One
can thus use this criterion to determine the minimal requirement on $M_\nu $
for the situation at hand.

As the finite-chain method leads to the greatest precision \cite{White}, let
us first look at some results using this algorithm. Table \ref{t1} lists a
few runs for a 25 atom chain. We have found that two sweeps are sufficient
in all situations. Further sweeping changes little to the numerical values.
Table \ref{t1} shows the various parameters used in each run: the oscillator
space dimension $M_\nu $, the number of block states kept $M_b$, the number
of target states $M_t$, and the estimated minimal error $\delta P$ arising
from the truncated oscillator space. We have calculated the numerical error
on the ground state correlation energy $\delta
E_{corr}=(E_{corr}^{DMRG}-E_{corr}^o)/E_{corr}^o$, where $E_{corr}=E_{gs}-%
\frac 12N_{mod}\hslash \omega _o$ and $E_{gs}$ is the ground state energy.
The superscripts ''$DMRG$'' and ''$o"$ refer to the computational and the
exact values, respectively. Note that $E_{gs}^o=\frac 12\sum_q\hbar \omega
_q $. This correlation energy is more significant than $E_{gs}$ since the
zero point energy of the local oscillators is not a numerically meaningful
quantity. We have also calculated the error on the correlation function
between the central atom and its first neighbor $\delta C=\left|
\left\langle (u_\ell -u_{\ell -1})^2\right\rangle _{DMRG}-\left\langle
(u_\ell -u_{\ell -1})^2\right\rangle _o\right| \allowbreak /\left\langle
(u_\ell -u_{\ell -1})^2\right\rangle _o$, the one on the oscillation
amplitude square of the central atom $\delta U=\left| \left\langle (u_\ell
)^2\right\rangle _{DMRG}-\left\langle (u_\ell )^2\right\rangle _o\right|
\allowbreak /\left\langle (u_\ell )^2\right\rangle _o$, and the error on the
average oscillator occupancy of the central atom $\delta n=\left|
\left\langle n\right\rangle _o-\left\langle n\right\rangle _{DMRG}\right|
/\left\langle n\right\rangle _o$. $\left\langle n\right\rangle
_o=\left\langle b_\ell ^{\dagger }b_\ell \right\rangle _o$, while $u_\ell $
and $b_\ell $ are defined in Eq. \ref{H3} and \ref{bd} respectively. The run
with 11 oscillator states and $M_b=10$ shows $\delta E_{corr}$ to be roughly
of the same size as $\delta P$. Increasing $M_\nu $ does not change $\delta
E_{corr}$ much. A saturation level has been reached for $M_\nu \geq 11$ even
though $\delta P$ keeps on decreasing. The central oscillator space is
sufficiently large that enlarging it any further does little. Why is this?
The reason is that important information contained in the central site is
not being relayed to the next iteration step because too few block states
are kept. Indeed, doubling $M_b$, as in the last run, considerably reduces
the error on the ground state energy. One notices that the error on the
nearest-neighbor correlation is of the same order as $\delta E_{corr}$. The
errors $\delta U$ and $\delta n$, however, are of the same order but much
larger than $\delta E_{corr}$. This is systematically found in all our
simulations. The reason is that these quantities sample the local oscillator
excitations to a much larger degree than the other quantities. The accuracy
on excitation energies is less than on the energy in the DMRG. Similar
findings, shown in Table \ref{t2}, were obtained for 49 and 99 atom chains.

In Table \ref{t3}, we used the infinite-chain algorithm to look at the
effect of increasing the number of target states. We see that the accuracy
gets better when using a few target states and worsens when using too many.
This is quite typical of the DMRG\cite{Moukouri}. Using four target states
transfers better information, through the projection procedure, to the block
states and improves $\delta n$ considerably. Using too many target states
dilutes the relevant information with irrelevant one. The runs with 30 block
states show that the infinite-chain algorithm with four target states can
perform as well as the finite-chain one (compare to the last entry in Table 
\ref{t1}) although the latter has just a single target state and a smaller
number of block states. One then has to weigh the numerical expenditure of
having more states as compared to doing two sweeps. This result is most
important as it shows how to make a proper use the infinite-chain algorithm.

We have shown that expressing the Hamiltonian in terms of local oscillators
leads to a short-range coupling between blocks and central atom(s). This is
a nice property to have when implementing the DMRG procedure. We have
observed that the numerical error introduced by the truncation of the local
oscillator space can be satisfactorily controlled by ab-initio selection of
the size of the oscillator space for the problem at hand. This applies to
any chain-length envisaged. We have found that the number of block states $%
M_b$ to be used is quite critical to the accuracy. As a rule of thumb, one
should keep at least something like 2 times the dimension of the local
oscillator space $M_\nu $. More important though, we have shown that the
infinite chain algorithm can be used in spite of the anticipated limitations
with regard to the lack of a fixed point and the logarithmic dependence of
the atomic motion as a function of chain length.

We have also found the usual DMRG truncation error $\delta E_{trunc}$ is
useless as an error indicator. We monitored it and observed it to be much
smaller than the error on the correlation energy $\delta E_{corr}$ or even $%
\delta P$. This is rather puzzling since, as seen in the previous section,
important errors stem from the quality of the projection of the target
states onto the block states. The reason has to do with the limited
information contained in the ground state of the super block with regard to
the central site. The central site of the super block is an inversion
symmetry center whereas it no longer is so at the end of the new block. The
symmetry breaking information that is required for a proper description of
the last site of the new block is thus contained in block states that have
little statistical weight in the density matrix.

The Hamiltonian we have studied is of an academic nature. By choosing the
local oscillator basis, we knowingly sacrificed its phonon conservation law
in order to obtain a form that is in harmony with the DMRG procedure. Our
study does, however, provide a necessary stepping stone for problems that do
not conserve the number of phonons. These are the more physical situations
like the Peierls or spin-Peierls chains or, quite generally, the
electron-phonon problem in one dimension. The computational expenditure
required for phonons is rather modest and would presumably remain so in
electron-phonon or spin-phonon problems with a gap as in \cite{PRL}.

We acknowledge the financial support of the Natural Sciences and Engineering
Research Council of Canada and the Fonds pour la Formation de Chercheurs et
d'Aide \`{a} la Recherche of the Qu\'{e}bec government.

\strut

\begin{table}[tbp] \centering%
{%
\caption{Finite-chain algorithm for 25 atoms. The parameters are: the
oscillator space dimension $M_\nu $, the number of block states $M_b$, the
number of target states $M_t$, the estimated minimal error on the oscillator
statistics $\delta P$, the numerical error on the ground state correlation
energy $\delta E_{corr}$, the one on the correlation function between the
central atom and its first neighbor $\delta C$, the one on the oscillation
amplitude squared of the central atom $\delta U$, and the error on the
average oscillator occupancy of the central atom $\delta n$.}\label{t1}} 
\begin{tabular}{llllllll}
$M_\nu $ & $M_b$ & $M_t$ & $\delta P$ & $\delta E_{corr}$ & $\delta C$ & $%
\delta U$ & $\delta n$ \\ \hline
11 & 10 & 1 & $10^{-5}$ & $1.9\times 10^{-5}$ &  &  & 0.013 \\ 
15 & 10 & 1 & $10^{-7}$ & $1.1\times 10^{-5}$ & $8.4\times 10^{-5}$ & $%
0.0032 $ & 0.006 \\ 
20 & 10 & 1 & $4\times 10^{-10}$ & $1.1\times 10^{-5}$ &  &  & $0.0059$ \\ 
15 & 20 & 1 & $10^{-7}$ & $2.8\times 10^{-6}$ & $2.6\times 10^{-6}$ & $%
1.7\times 10^{-4}$ & $3.6\times 10^{-4}$%
\end{tabular}
\end{table}%

\begin{table}[tbp] \centering%
{%
\caption{ Finite-chain algorithm for chains of $N_a$=49 and 99 atoms. The
parameters are defined in Table I.}\label{t2}} 
\begin{tabular}{lllllllll}
$M_\nu $ & $M_b$ & $M_t$ & $N_a$ & $\delta P$ & $\delta E_{corr}$ & $\delta
C $ & $\delta U$ & $\delta n$ \\ \hline
15 & 10 & 1 & 49 & $3\times 10^{-6}$ & $4.4\times 10^{-4}$ & $2.1\times
10^{-4}$ & $0.026$ & 0.05 \\ 
15 & 20 & 1 & 49 & $3\times 10^{-6}$ & $2.5\times 10^{-5}$ & $1.6\times
10^{-5}$ & $0.0032$ & 0.006 \\ 
20 & 20 & 1 & 49 & $10^{-8}$ & $1.3\times 10^{-5}$ & $1.2\times 10^{-5}$ & $%
0.0014$ & 0.0026 \\ 
20 & 30 & 1 & 49 & $10^{-8}$ & $2.1\times 10^{-6}$ & $8.8\times 10^{-7}$ & $%
2.2\times 10^{-4}$ & $4.0\times 10^{-4}$ \\ 
15 & 20 & 1 & 99 & $3\times 10^{-5}$ & $9.8\times 10^{-5}$ & $4.7\times
10^{-5}$ & $0.028$ & 0.045 \\ 
20 & 30 & 1 & 99 & $3\times 10^{-7}$ & $1.1\times 10^{-5}$ & $3.2\times
10^{-6}$ & 0.0038 & 0.0063
\end{tabular}
\end{table}%

\begin{table}[tbp] \centering%
{%
\caption
{Infinite chain algorithm for 25 atoms. The parameters are defined in Table I.}
\label{t3}} 
\begin{tabular}{lllll}
$M_\nu $ & $M_b$ & $M_t$ & $\delta E_{corr}$ & $\delta n$ \\ \hline
15 & 20 & 1 & $3.7\times 10^{-4}$ & $0.0035$ \\ 
15 & 20 & 4 & $2.8\times 10^{-4}$ & 0.0015 \\ 
15 & 20 & 7 & $1.7\times 10^{-3}$ & 0.0085 \\ 
15 & 30 & 4 & $2.6\times 10^{-5}$ & $3.5\times 10^{-4}$ \\ 
15 & 30 & 1 & $1.5\times 10^{-4}$ & .00145
\end{tabular}
\end{table}%

\end{document}